\documentclass[preprint,12pt]{elsarticle}%
\usepackage{amssymb}
\usepackage{amsfonts}
\usepackage{amsmath}
\usepackage{graphicx}%
\providecommand{\U}[1]{\protect\rule{.1in}{.1in}}
\graphicspath{{E:/dottorato/Coulomb_damper/Testo/Paper/Immagini/}}
\journal{journal}

\begin{document}
%
\begin{frontmatter}%


%

\title
{On the interaction of viscoelasticity and waviness in enhancing the pull-off force in sphere/flat contacts}%

%

\author{M. Ciavarella}%
%

\address
{Politecnico di BARI. DMMM department. Viale Gentile 182, 70126 Bari. Mciava@poliba.it}%
%

\begin{abstract}%

Motivated by roughness-induced adhesion enhancement (toughening and
strengthening) in low modulus materials, we study the detachment of a sphere
from a substrate in the presence of both viscoelastic dissipation at the
contact edge, and roughness in the form of a single axisymmetric waviness. We
show that the roughness-induced enhancement found by Guduru and coworkers for
the elastic case (i.e. at very small detachment speeds) tends to disappear
with increasing speeds, where the viscoelastic effect dominates and the
problem approaches that of a smooth sphere. This is in qualitative agreement
with the original experiments of Guduru's group with gelatin. The cross-over
velocity is where the two separate effects are comparable. Viscoelasticity
effectively damps roughness-induced elastic instabilities, and make their
effects much less important.%

\end{abstract}%
%

%

\end{frontmatter}%



\section{\bigskip Introduction}

It is well known that adhesion of hard solids is difficult to measure at
macroscopic scales, and Fuller and Tabor (1975) proved that even in low
modulus materials (they used rubbers with $E\sim1$ $\mathrm{MPa}$), a
$\sim1\left[  \mu m\right]  $ of roughness destroy adhesion almost completely,
despite van der Waals adhesive forces are quite strong, giving the so called
"adhesion paradox" (Kendall, 1975). Adhesion of macroscopic bulk objects
requires smooth surfaces, and at least one of the solids has to have a very
low elastic modulus. Dahlquist (1969a, 1969b) while working at 3M proposed a
criterion largely used in the world of adhesives, namely that the elastic
Young modulus should be smaller than $\sim1$ $\mathrm{MPa}$ to achieve
stickiness even in the presence of roughness. This clearly is just a rough
indication, but Tiwari \textit{et al}.(2017) find for example that the work of
adhesion (at a given retraction speed) is reduced of a factor 700 for a rubber
in contact with a rough hard sphere when the rubber modulus is $E=2.3MPa$, but
is actually \textit{increased} because of roughness by a factor 2 when the
rubber has modulus $E=0.02MPa$. The threshold doesn't change much even if we
consider nanometer scale roughness, as in the recent results of Dalvi
\textit{et al.} (2019) for pull-off of PDMS hemispheres having four different
elastic moduli against different roughened plates: Dahlquist's criterion seems
to work surprisingly well, as while there is little effect of roughness for
the 3 cases of low modulus up to near $E=2\left[  MPa\right]  $, roughness has
strong effect both during approach and retraction for the high modulus
material ($E=10\left[  MPa\right]  )$, where the hysteresis may be due partly
to viscoelastic effects\footnote{Despite the authors intended to remove as
much as possible rate-dependent effects by applying a retraction rate of only
$60nm/s$). The authors claim a good correlation of the energy loss during the
cycle of loading and withdrawing with the product of the real contact area at
maximum preload with the "intrinsic" work of adhesion. Notice that this would
not work for a smooth sphere where JKR theory predicts that the energy loss is
independent on preload and indeed the data of Dalvi et al. (2019) with the
lowest roughness do show almost a constant trend. Also, the hard material case
shows almost no energy loss.}. However, for the 3 low modulus materials,
roughness almost \textit{systematically increases} the work of adhesion rather
than decreasing it as for the high modulus material, for a given retraction speed.

Roughness-induced adhesion enhancement was measured with some surprise first
by Briggs \& Briscoe (1977), and Fuller \& Roberts (1981), and
Persson-Tosatti's (2001) theory above attributes it to the increase of surface
area induced by roughness\footnote{Whereas adhesion reduction is attributed by
Persson and Tosatti (2001) to the elastic energy to flatten roughness, which
is proportional to the elastic modulus.}. Another mechanism was put forward by
Guduru and collaborators (Guduru, 2007, Guduru and Bull, 2007). Guduru
considered a spherical contact having a concentric axisymmetric waviness, and
considers the contact is complete over the contact area. The waviness gives
rise to oscillations in the load-approach curve which result in up to factor
20 increase of the pull-off with respect to the standard smooth sphere case of
the JKR theory (Johnson, Kendall \&\ Roberts, 1971). Also, the curves fold on
each other so that we expect jumps at some points in the equilibrium curve.
which corresponds to dissipation and emission of elastic waves in the
material, and results in strong hysteresis. Later, Kesari and Lew (2011)
noticed that Guduru's solution has an elegant "envelope" obtained by expanding
asymptotically for very small wavelength of the waviness.

But most soft materials are viscoelastic, and therefore there is a strong
velocity dependence of the pull-off result. Many authors (Gent and Schultz,
1972, Barquins and Maugis 1981, Gent, 1996, Gent \& Petrich 1969, Andrews \&
Kinloch, 1974, Barber \textit{et al}, 1989, Greenwood \& Johnson, 1981, Maugis
\& Barquins, 1980, Persson \& Brener, 2005) have proposed that the process of
peeling involves an effective work of adhesion $w$ which is the product of the
thermodynamic (Dupr\'{e}) work of adhesion $w_{0}$ and a function of velocity
of peeling of the contact/crack line and temperature, as long as there is no
bulk viscoelasticity involved, over a large range of crack speeds, namely of
the form that has been validated also by a large amount of data including
peeling tests at various peel angles%
\begin{equation}
w=w_{0}\left[  1+k\left(  a_{T}v_{p}\right)  ^{n}\right]  \label{wvisco}%
\end{equation}
where $k,n$ are constants of the material, with $n$ in the range $0.1-0.8$ and
$a_{T}$ is the WLF factor (Williams, Landel \& Ferry, 1955) which permits to
translate results at various temperatures $T$ from measurement at a certain
standard temperature. This form of effective work of adhesion was obtained
also from theoretical models using either Barenblatt models or crack tip
blunting models (Barber Donley and Langer 1989, Greenwood and Johnson, 1981,
Persson \&\ Brener, 2005) generally using simple a single relaxation time
model for the materials. Actually, Persson \&\ Brener (2005) showed that for a
frequency dependent viscoelastic modulus $E\left(  \omega\right)  \sim
\omega^{1-s}$, \ $0<s<1$, in the transition region between the "rubbery
region" and the "glassy region" (where the strong internal damping occurs
important for energy loss processes), the equation (\ref{wvisco}) is satisfied
at intermediate velocities with $n=\left(  1-s\right)  /\left(  2-s\right)  $
(so that $0<n<1/2,$ in agreement with most of the range cited above).\ There
may be deviations for materials having a complex behaviour with many
relaxation times or for considering the role of finite size of the system or
of very high temperature at the crack tip which depends on thermal diffusivity
of the material, see recent review (Rodriguez \textit{et al}., 2020).

The effective "toughness" $w$ can increase of various orders of magnitude over
$w_{0}$ as the velocity increases (more precisely, of the ratio $E\left(
\infty\right)  /E\left(  0\right)  $, where $E\left(  \omega\right)  $ is the
frequency dependent elastic modulus), and the pull-off of a sphere has also
been effectively measured to increase of various orders of magnitude over an
increase of peeling speed (Barquins \&\ Maugis, 1981). On the contrary, during
crack closure the effective work of adhesion is even smaller than $w_{0}$,
(this time it is reduced by the ratio $E\left(  0\right)  /E\left(
\infty\right)  $, see Greenwood and Johnson, 1981), so in some cases loading
could become essentially an elastic model without adhesion.

Equation (\ref{wvisco}) generalizes the thermodynamic equilibrium of elastic
cracks for the strain energy release $G$: namely, it provides a condition for
crack edge velocity --- when $G>w_{0}$, the crack accelerate under the force
$G-w_{0}$ applied per unit length of crack, until a limit speed $v_{p}$ for
equilibrium is found, depending on the loading conditions. For example,
$G-w_{0}$ is a constant for classical peeling experiments, whereas it
monotonically increases for flat punches, and has a much richer behaviour for
the smooth sphere. Therefore, for imposed tensile load smaller in absolute
value than the JKR pull-off value $P_{0}=3/2\pi wR$, the contact area simply
decreases to another equilibrium value (given asymptotically by JKR theory),
while for imposed load below the JKR value, it decreases with non monotonic
velocity but without the JKR pull-off instability, so up to complete
detachment. Therefore, pull-off depends on the loading condition: can be
anything greater than $P_{0}$ if load is imposed, whereas it is a precise
function of the retraction rate in an experiment where the cross-head of a
rigid machine keeps the remote approach velocity as constant.

\bigskip Various authors (Barquins \& Maugis, 1981, Greenwood \&\ Johnson,
1981, Muller, 1999) have studied the peeling of viscoelastic spheres with the
above form of fracture mechanics formulation (\ref{wvisco}), and some
approximate scaling results have also been given (Muller, 1999), but a
theoretical or numerical investigation about the coupled effect of
viscoelasticity and roughness has not been attempted, in the best of the
author's knowledge, not even with numerical simulations. It seems that in
general viscoelasticity can only increase the "tack" i.e. the force or the
work needed to detach two solids, whereas the role of roughness is more
controversial, as we have discussed above. We are aware of the complexity of
the general problem, so here, we tackle the study of a simple problem, that of
a sphere with a single wave of roughness, which generalizes the relatively
recent work of Guduru and collaborators (Guduru, 2007, Guduru and Bull, 2007)
and following related literature, to the case of a viscoelastic substrate.

\section{The theory}

\bigskip We consider the Guduru contact problem for a sphere against a flat
surface, where the gap is defined as $f\left(  r\right)  =\frac{r^{2}}%
{2R}+A\left(  1-\cos\frac{2\pi r}{\lambda}\right)  $, where $R$ is the sphere
radius, $\lambda$ is wavelength of roughness, $A$ is its amplitude.

\begin{center}
\bigskip%
\begin{tabular}
[c]{l}%
\centering\includegraphics[height=65mm]{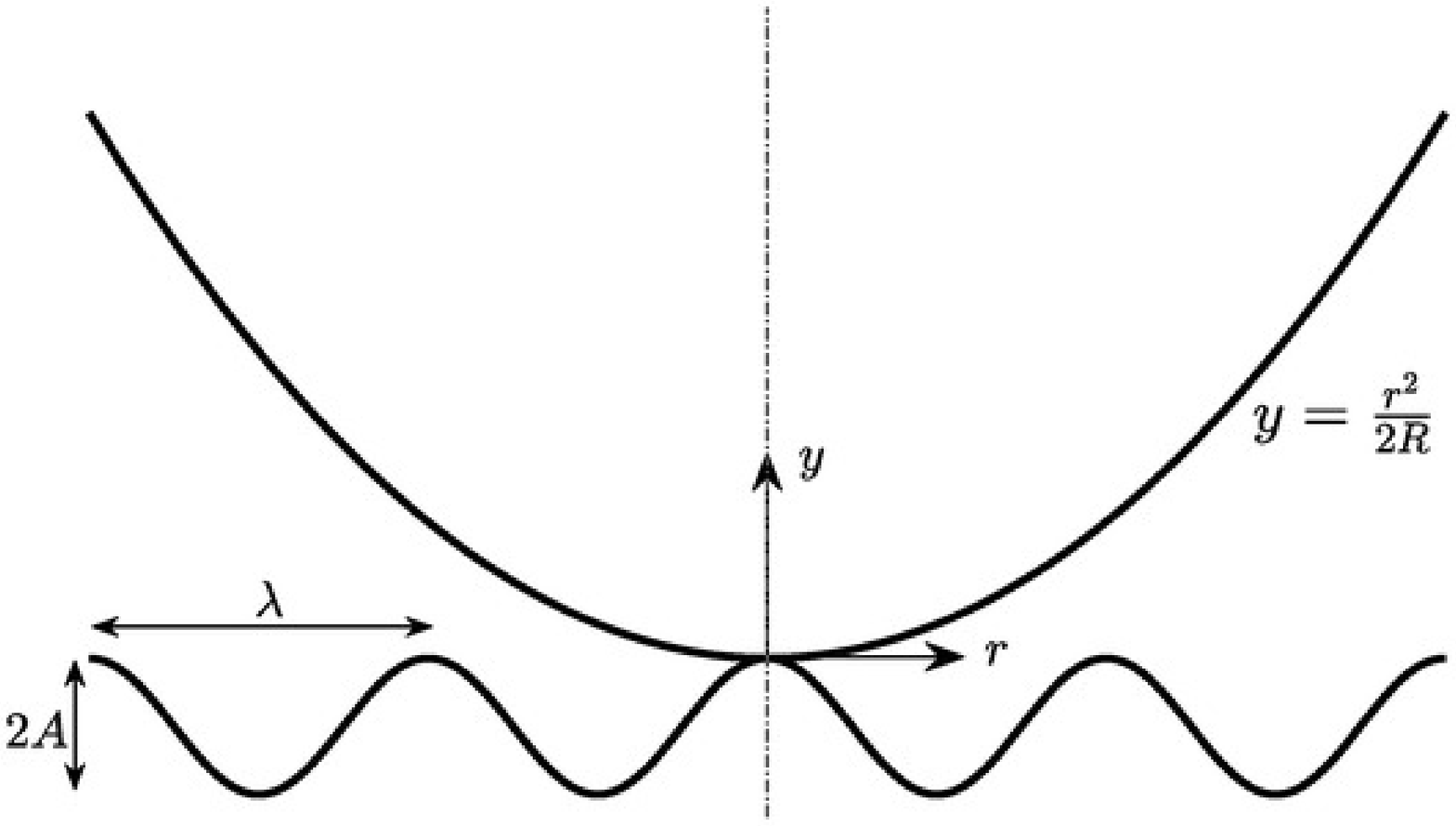}
\end{tabular}

Fig.1 - The geometry of the problem. A sphere of radius $R$ with a simple
roughness being a single axisymmetric wave with wavelength $\lambda$ and
amplitude $A$.

\end{center}

The Guduru problem can be solved by considering the stress intensity factor
$K$ at the contact edge (radius $r=a$) or equivalently the strain energy
release rate $G$ (Guduru, 2007)%
\begin{equation}
G\left(  a,P\right)  =\frac{K\left(  a,P\right)  ^{2}}{2E^{\ast}}%
=\frac{\left(  P_{1}\left(  a\right)  -P\right)  ^{2}}{8\pi E^{\ast}a^{3}}
\label{G}%
\end{equation}
where $E^{\ast}=E/\left(  1-\nu^{2}\right)  $ is plane strain elastic modulus
(i.e. $E$ is Young's modulus and $\nu=0.5$ Poisson ratio generally equal to
0.5 in rubbery materials, while we consider the countersurface is generally
much more rigid so we neglect its elastic properties).

Here, $P_{1}\left(  a\right)  $ is the load required to maintain a contact
radius $a$ in the absence of adhesion, while $P$ is the smaller load to
maintain the same contact radius in the presence of adhesion. In particular
standard contact mechanics gives (Guduru, 2007)
\begin{equation}
P_{1}\left(  a\right)  =2E^{\ast}\left\{  \left(  \frac{2}{R}+\frac{4\pi^{2}%
A}{\lambda^{2}}\right)  \frac{a^{3}}{3}+\frac{\pi Aa}{2}H_{1}\left(
\frac{2\pi a}{\lambda}\right)  -\frac{\pi^{2}Aa^{2}}{\lambda}H_{2}\left(
\frac{2\pi a}{\lambda}\right)  \right\}  \label{P1}%
\end{equation}
where $H_{n}$ are Struve functions of order $n$.

In the adhesionless conditions, the remote approach (positive for compression)
is
\begin{equation}
\alpha_{1}\left(  a\right)  =\frac{a^{2}}{R}+\pi^{2}\frac{A}{\lambda}%
aH_{0}\left(  \frac{2\pi a}{\lambda}\right)  \label{alfa1}%
\end{equation}
so in the adhesive condition we have to decrease this by an amount given by a
flat punch displacement giving the general result for approach
\begin{equation}
\alpha\left(  a,P\right)  =\frac{a^{2}}{R}+\pi^{2}\frac{A}{\lambda}%
aH_{0}\left(  \frac{2\pi a}{\lambda}\right)  -\frac{P_{1}\left(  a\right)
-P}{2E^{\ast}a} \label{alfa}%
\end{equation}

From (\ref{alfa}), we can obtain the general equation for the load as a
function of contact radius and approach
\begin{equation}
P\left(  a,\alpha\right)  =P_{1}\left(  a\right)  +2E^{\ast}a\alpha\left(
a\right)  -2E^{\ast}\frac{a^{3}}{R}-\pi^{2}\frac{2E^{\ast}A}{\lambda}%
a^{2}H_{0}\left(  \frac{2\pi a}{\lambda}\right)  \label{P}%
\end{equation}
where $P_{1}\left(  a\right)  $ is given by (\ref{P1}) above. Imposing the
condition of thermodynamic equilibrium $G\left(  a\right)  =w_{0}$ using
(\ref{G}) and (\ref{P1}) permits to write the Guduru solution explicitely as
parametric equations of the contact radius $a$
\begin{align}
P\left(  a\right)   &  =P_{1}\left(  a\right)  -a^{3/2}\sqrt{8\pi w_{0}%
E^{\ast}}\label{PGuduru}\\
\alpha\left(  a\right)   &  =\alpha_{1}\left(  a\right)  -a^{1/2}\sqrt{2\pi
w_{0}/E^{\ast}} \label{alfaGuduru}%
\end{align}

Using the Kesari \&\ Lew (2011) expansion, Ciavarella (2016) obtained that the
Guduru solution has oscillations bounded between two \textit{exact} JKR
(Johnson, Kendall \&\ Roberts, 1971) \textit{envelope curves} for the smooth
sphere, but with a corrected (\textit{enhanced} or \textit{reduced},
respectively for unloading or loading) surface energy
\begin{align}
P_{env}\left(  a\right)   &  =\frac{4}{3R}E^{\ast}a^{3}-a^{3/2}\sqrt{8\pi
wE^{\ast}}\left(  1\pm\frac{1}{\sqrt{\pi}\alpha_{KLJ}}\right) \label{Penv}\\
\alpha_{env}\left(  a\right)   &  =\frac{a^{2}}{R}-a^{1/2}\sqrt{\frac{2\pi
w}{E^{\ast}}}\left(  1\pm\frac{1}{\sqrt{\pi}\alpha_{KLJ}}\right)
\label{alfaenv}%
\end{align}
where
\begin{equation}
\alpha_{KLJ}=\sqrt{\frac{2w_{0}\lambda}{\pi^{2}E^{\ast}A^{2}}} \label{alfaKLJ}%
\end{equation}
is the parameter Johnson (1995) introduced for the JKR adhesion of a nominally
flat contact having a single scale sinusoidal waviness of amplitude $A$ and
wavelength $\lambda$. Thus, since (\ref{Penv},\ref{alfaenv}) are JKR equation
for a smooth sphere of radius $R$, the factor
\begin{equation}
\frac{w_{eff}}{w_{0}}=\left(  1+\frac{1}{\sqrt{\pi}\alpha_{KLJ}}\right)  ^{2}
\label{enhancement}%
\end{equation}
is an roughness-induced increase which holds as long as a compact contact area
can be obtained, which requires not too large roughness, and/or sufficiently
strong precompression. In practice, factors up to 20 have been obtained also
experimentally by Guduru \&\ Bull (2007), although of course these were
achieved in geometry built for the specific goal to achieve very large
enhancement. Fig.2 elucidates the behaviour of the oscillations in the Guduru
solution for a representative case, which we shall later extend to the
viscoelastic solution. Given these gulfs and reentrances, in the elastic
solution, the real followed path will depend on the loading condition. For a
soft system (close to "load control"), there will be horizontal jumps in
approach while in a stiff system (close to "displacement control") there will
be vertical jumps to the next available stable position. In both cases there
will be areas "neglected" during these jumps which represent mechanical
dissipated energy. Indeed, in the "envelope" solution of
Kesari-Lew-Ciavarella, the combined effect of these jumps results in the
different JKR loading and unloading curves which give an additional hysteresis
with respect to the standard JKR case, where the only hysteresis comes a
single elastic instability in pull-in and another (different) single
instability at pull-off. The dashed lines in Fig.2 are the
Kesari-Lew-Ciavarella envelopes \ (\ref{Penv},\ref{alfaenv}) using
(\ref{alfaKLJ}).

\begin{center}%
\begin{tabular}
[c]{l}%
\centering\includegraphics[height=65mm]{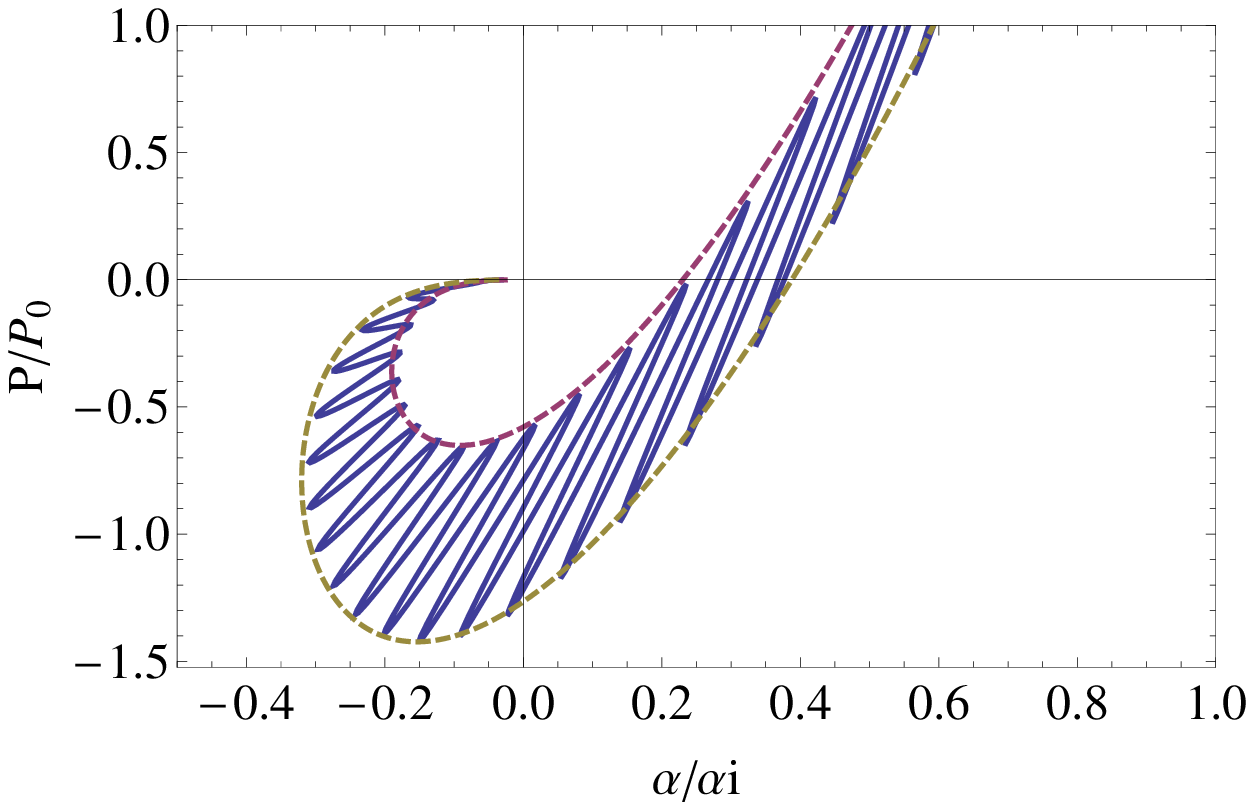}
\end{tabular}

Fig.2 - The load-approach curve in the Guduru elastic problem with $E^{\ast
}=16500\left[  Pa\right]  $; $R=0.23\left[  m\right]  $; $w_{0}=0.008\left[
J/m^{2}\right]  $; $\lambda/R=0.002$; $A/\lambda=0.005$, and reference $a_{i}$
$=0.01\left[  m\right]  $.
\end{center}

\subsection{\bigskip Viscoelastic problem}

For a given remote applied withdrawing of the sphere $v=-\frac{d\alpha}{dt}$,
we can write the velocity of the contact edge as
\begin{equation}
v_{p}=-\frac{da}{dt}=v\frac{da}{d\alpha} \label{v}%
\end{equation}
The condition $G\left(  a\right)  =w$ (which replaced the thermodynamic
equilibrium $G\left(  a\right)  =w_{0}$ for the elastic sphere) therefore
defines a differential equation for $a=a\left(  \alpha\right)  $, obtained
using (\ref{G},\ref{wvisco},\ref{v})
\begin{equation}
\frac{1}{k^{1/n}a_{T}v}\left(  \frac{\left(  P_{1}\left(  a\right)  -P\right)
^{2}}{8\pi E^{\ast}a^{3}w_{0}}-1\right)  ^{1/n}=\frac{da}{d\alpha} \label{1}%
\end{equation}

Hence, using (\ref{P}) and defining dimensionless parameters
\begin{equation}
V=k^{1/n}a_{T}v\qquad;\qquad\zeta=\left(  \frac{2\pi w_{0}}{RE^{\ast}}\right)
^{1/3} \label{dimensionless}%
\end{equation}
we write (\ref{1}) as%
\begin{equation}
\frac{da}{d\alpha}=\frac{1}{V}\left[  \frac{\left(  R/a\right)  }{\zeta^{3}%
}\left(  \frac{\alpha}{R}-\frac{a^{2}}{R^{2}}-\pi^{2}\frac{A}{\lambda}\frac
{a}{R}H_{0}\left(  \frac{2\pi a}{\lambda}\right)  \right)  ^{2}-1\right]
^{1/n}%
\end{equation}
which can be solved with a numerical method. After a solution is obtained for
$a=a\left(  \alpha\right)  $, we substitute back into (\ref{P}) to compute the
load. Notice that, for a given starting point of the peeling process in terms
of load $P$, the term under parenthesis in (\ref{1}) is zero, and hence
$\frac{da}{d\alpha}$ starts off zero giving some delay with respect to the
elastic curve, which is hard to eliminate even at very low withdrawal speeds.

\section{Results}

\subsection{Rough sphere}

We consider first a viscoelastic material having $n=0.33$, and dimensionless
withdrawal velocity $V=0.0002,0.002,0.02,0.2,2$; the other constants, as
indicated in Fig.1, are $E^{\ast}=16500\left[  Pa\right]  $; $R=0.23\left[
m\right]  $; $w_{0}=0.008\left[  J/m^{2}\right]  $; $\lambda/R=0.002$;
$A/\lambda=0.005$. This corresponds to an "adiabatic" \textit{elastic}
enhancement of the pull off according to the equation derived from the Guduru
theory (\ref{enhancement}) of $w_{eff}/w_{0}=1.42$. We indicate with
$P_{0}=3/2\pi w_{0}R$ the JKR value of pull-off for the smooth sphere. The
loading curve follows the elastic solution (\ref{PGuduru}, \ref{alfaGuduru}),
and we start withdrawing the indenter from a reference value of $a_{i}$
$=0.01\left[  m\right]  $. Corresponding values of initial approach
$\alpha_{i}$ and load $P_{i}$ can therefore found from (\ref{PGuduru},
\ref{alfaGuduru}). Numerical solutions are found with the NDSolve algorithm in
Mathematica with default options. Fig.3a shows the obtained load-approach
curve in terms of $P/P_{0}$ and $\alpha/\alpha_{i}$ where $P_{0}$ is JKR
pull-off of the smooth sphere, and Fig.3b shows the contact radius $a/a_{i}$
peeling as a function of approach. The inner black wavy curves are the
equilibrium Guduru solutions, and the other 5 curves are obtained numerically
for increasing dimensionless velocities of withdrawal
$V=0.0002,0.002,0.02,0.2,2$.\ As expected, the viscoelastic peeling terminates
only when contact radius is zero, and not \ at the JKR\ unstable radius.
However, the minimum load is found for a contact radius which, for low
velocities, is not too different from the unstable pull-off contact radius in
JKR theory.

\begin{center}%
\begin{tabular}
[c]{l}%
\centering\includegraphics[height=65mm]{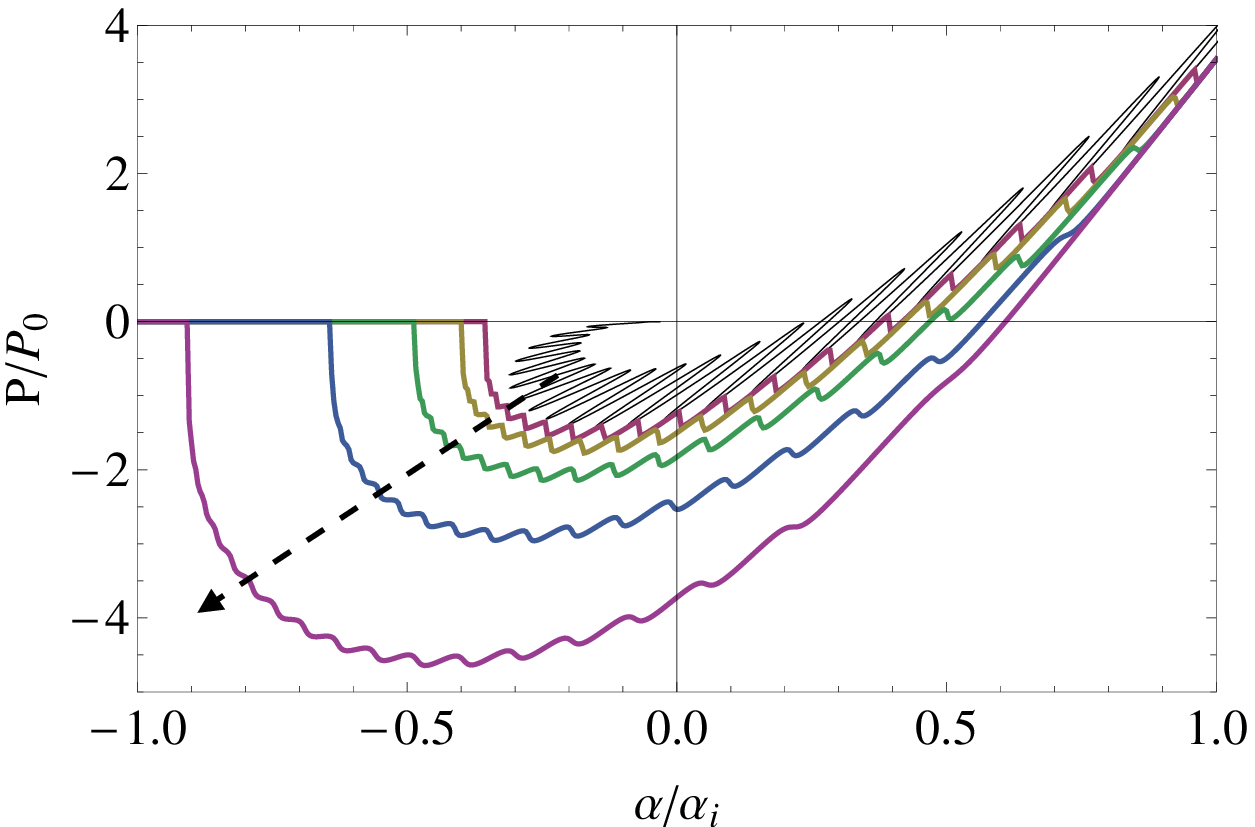}(a)
\end{tabular}

\begin{tabular}
[c]{l}%
\centering\includegraphics[height=65mm]{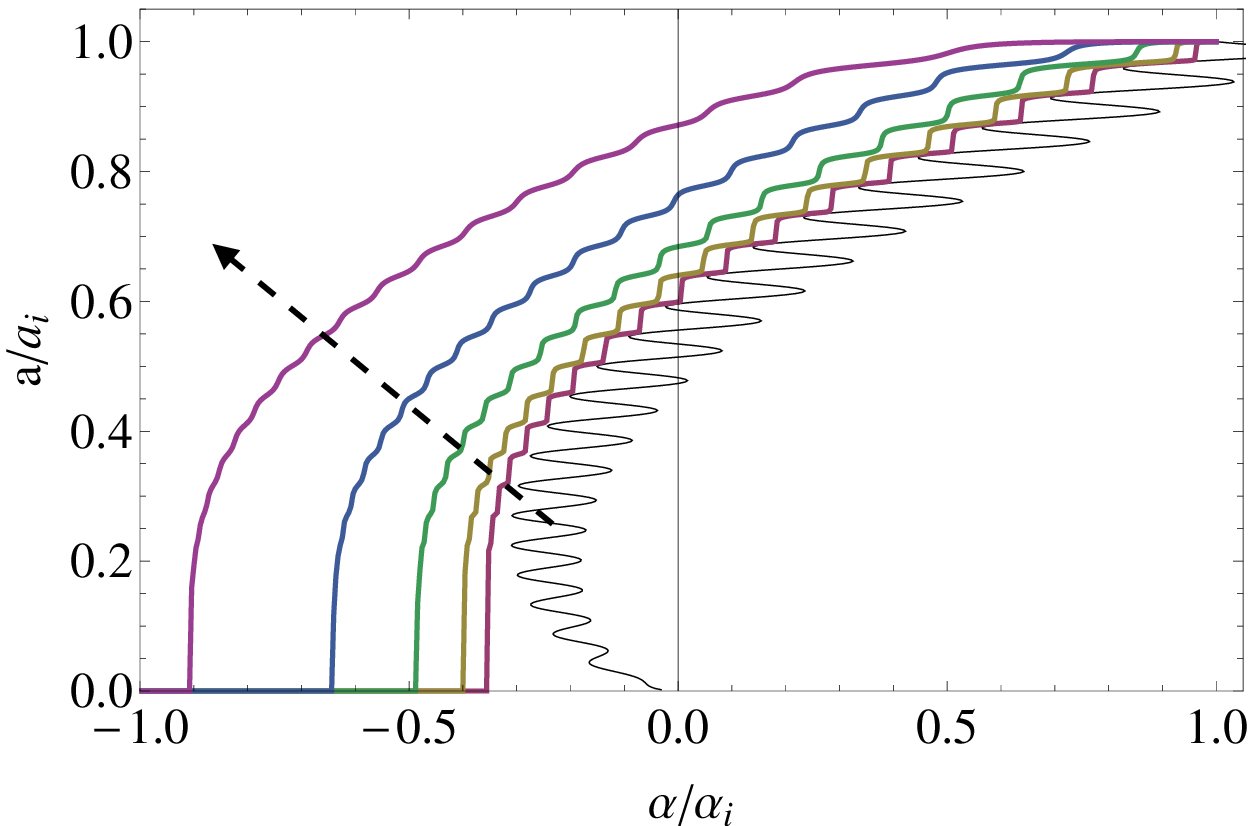}(b)
\end{tabular}

Fig.3 - Load $P/P_{0}$ (a) and the contact radius $a/a_{i}$ (b) as a function
of approach $\alpha/\alpha_{i}$. $P_{0}$ is JKR pull-off of the smooth sphere,
and $\alpha_{i}$ the initial value of approach for unloading. The inner black
wavy curve is the equilibrium Guduru solution, and the other 5 curves
increasingly departing from it are obtained numerically for increasing
dimensionless velocities of withdrawal (see the verse of the arrow)
$V=0.0002,0.002,0.02,0.2,2$. Here, $n=0.33$, and other constants as indicated
in Fig.1.
\end{center}

Fig.4a,b,c give some detail of the solution at the lowest dimensionless
velocity of withdrawal $V=0.0002$. In particular, Fig.4a shows clearly that
the numerical solution follows closely the prediction of the Guduru elastic
solution under displacement control, as expected, with almost sharp jumps of
the force at specific values of the approach. After the jump, the solution
seems to return to the Guduru equilibrium solution. Obviously with the
viscoelastic theory, the strict elastic solution should be obtained
asymptotically at extremely low velocities, but the differential equation
would then become very "stiff" corresponding to numerical difficulties
following the jumps. The same behaviour is clarified in terms of the contact
radius in Fig.4b which follows very closely the Guduru solution in some
periods of time, then extends a little before jumping almost abruptly to the
following branch of the equilibrium curve. In other words, the curve does not
have a "rainflow" type of behaviour over the Guduru equilibrium solution,
which would be the elastic real behaviour with jump-instabilities, but the
contact radius "drops" over the Guduru curve only after some delays. This is
further clarified in Fig.4c, where the velocity of the contact line
$da/d\alpha=v_{p}/v$ is found to follow an oscillatory trend with "bursts" of
very high (but finite) velocity where the peeling velocity is a much larger
value than the imposed withdrawing velocity, after which the velocity drops to
a low value which is where the contact area approaches the adiabatic Guduru
curve since $G\simeq w_{0}$, and which increases progressively with the
decreasing approach. Slowly, the solution departs from the Guduru elastic one,
because of the cumulative effects of the acceleration periods. However, from
Fig.3, and Fig.4d, for high velocities, we see that there are no real "jumps",
and the solution curve is generally smoother, with the difference between the
slow regime and the fast regime being smaller. Also notice in particular that
while the velocity of peeling remains in every case equal to zero at the
initial point, it remains closer to zero for a much extensive range of
approach for high velocities, resulting in a curve departing away from the
equilibrium Guduru curve immediately. This effect at high velocity produces
curves that are generally closer to the viscoelastic curves for the smooth
sphere, and therefore closer results for pull-off and work for pull-off. To
see this from a quantitative point of view, since even the smooth sphere
problem requires a numerical solution, we describe some results in the next paragraph.

\begin{center}
$%
\begin{array}
[c]{cc}%
\begin{tabular}
[c]{l}%
\centering\includegraphics[height=45mm]{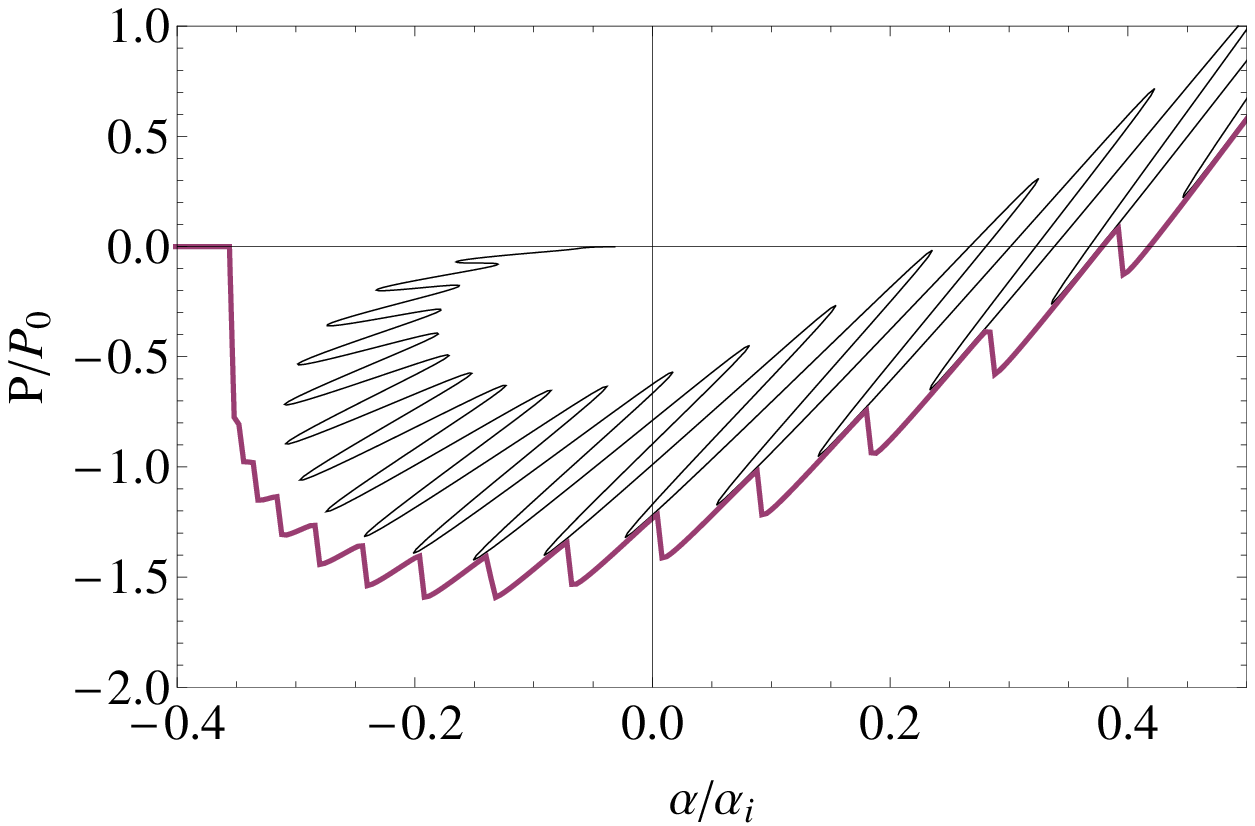}
\end{tabular}
&
\begin{tabular}
[c]{l}%
\centering\includegraphics[height=45mm]{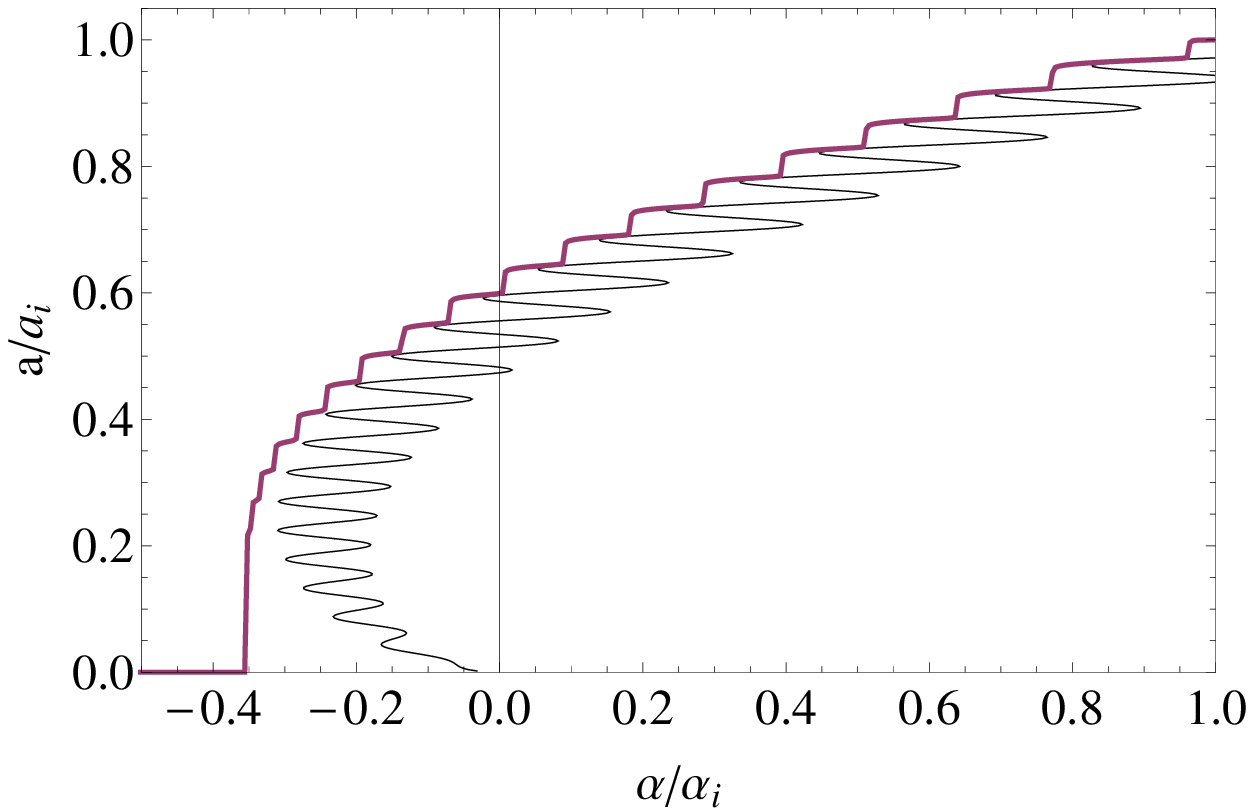}
\end{tabular}
\\
\text{(a)} & \text{(b)}%
\end{array}
$

$%
\begin{array}
[c]{cc}%
\begin{tabular}
[c]{l}%
\centering\includegraphics[height=45mm]{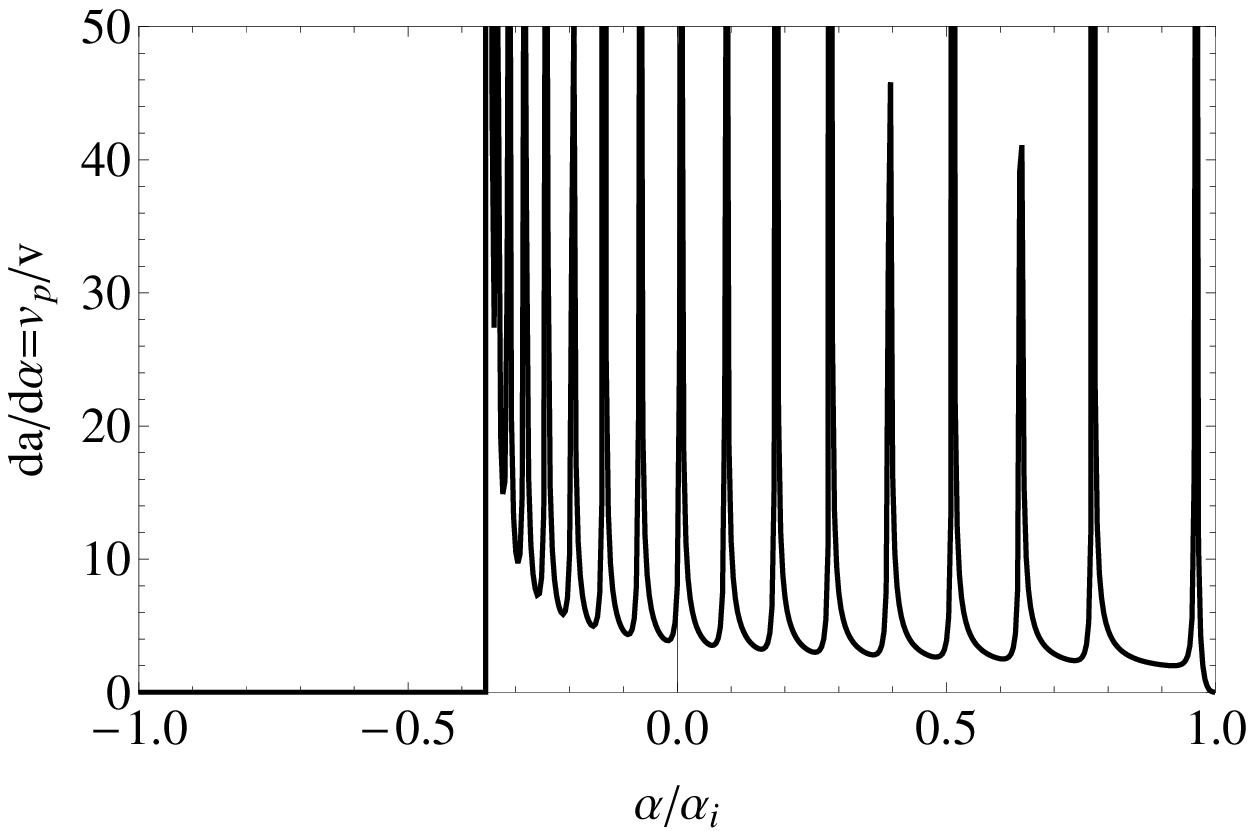}
\end{tabular}
&
\begin{tabular}
[c]{l}%
\centering\includegraphics[height=45mm]{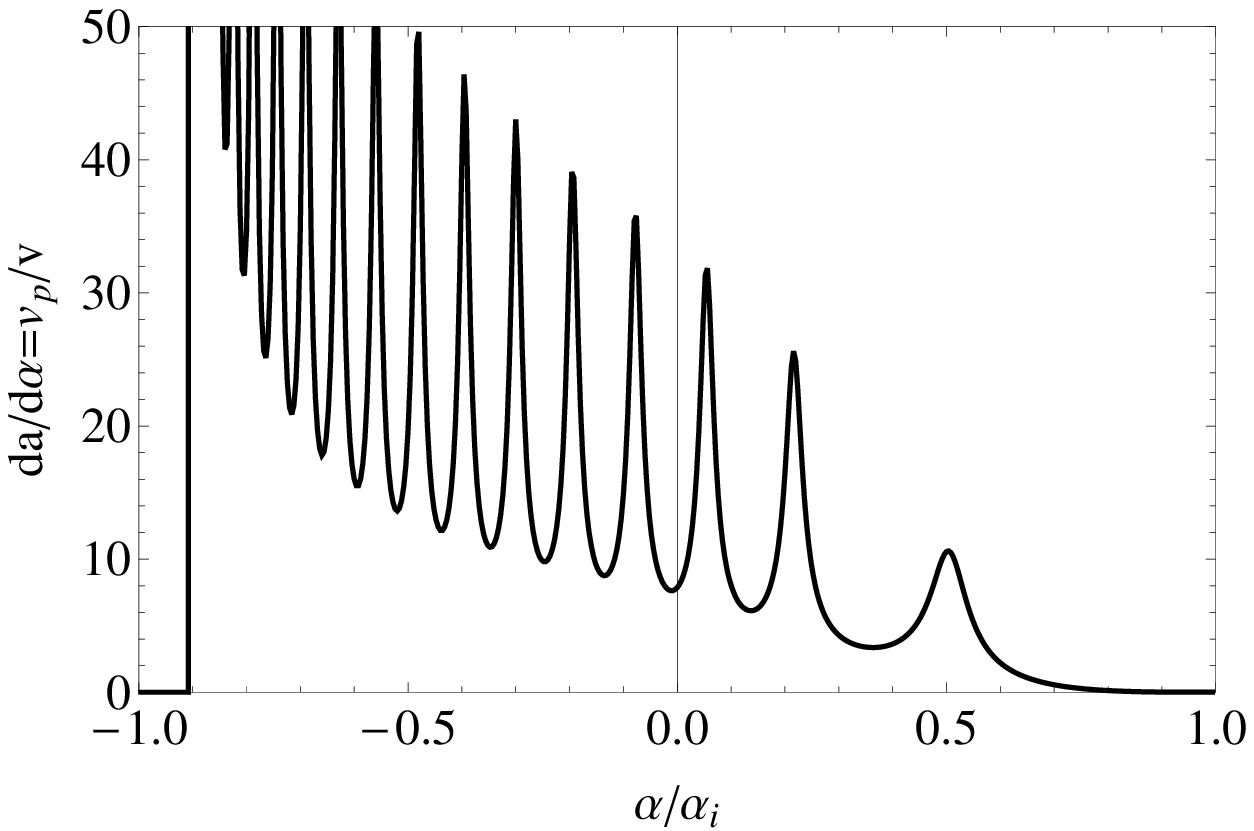}
\end{tabular}
\\
\text{(c)} & \text{(d)}%
\end{array}
$

Fig.4 - Detail of the solution at the lowest dimensionless velocity of
withdrawal (a,b,c) $V=0.0002,$ and (d) $V=2$. Here, $n=0.33$, and other
constants as indicated in Fig.1. In particular (a) Load-approach (b) contact
radius vs approach (c) velocity of contact line $da/d\alpha=v_{p}/v$. (d)
velocity of contact line $da/d\alpha=v_{p}/v$ but for the highest
dimensionless speed $V=2$. \ 
\end{center}

\subsection{Smooth sphere}

We solve the problem for the smooth sphere under the same set of conditions
(the equation for the smooth sphere are obviously obtained for $A=0$ above),
and obtained smooth results are shown in Fig.5. Notice that as we discussed in
the theory paragraph, in the initial point $\frac{da}{d\alpha}=0$, and we find
again (Fig.5c) that the velocity remains practically zero for a longer time
when $V$ is bigger.\ Here, contrary to the rough sphere, but also contrary to
the smooth sphere in the case when load is constant, the velocity of the
contact line increases monotonically from zero to infinite when pull-off
occurs at zero contact area.

\begin{center}%
\begin{tabular}
[c]{l}%
\centering\includegraphics[height=45mm]{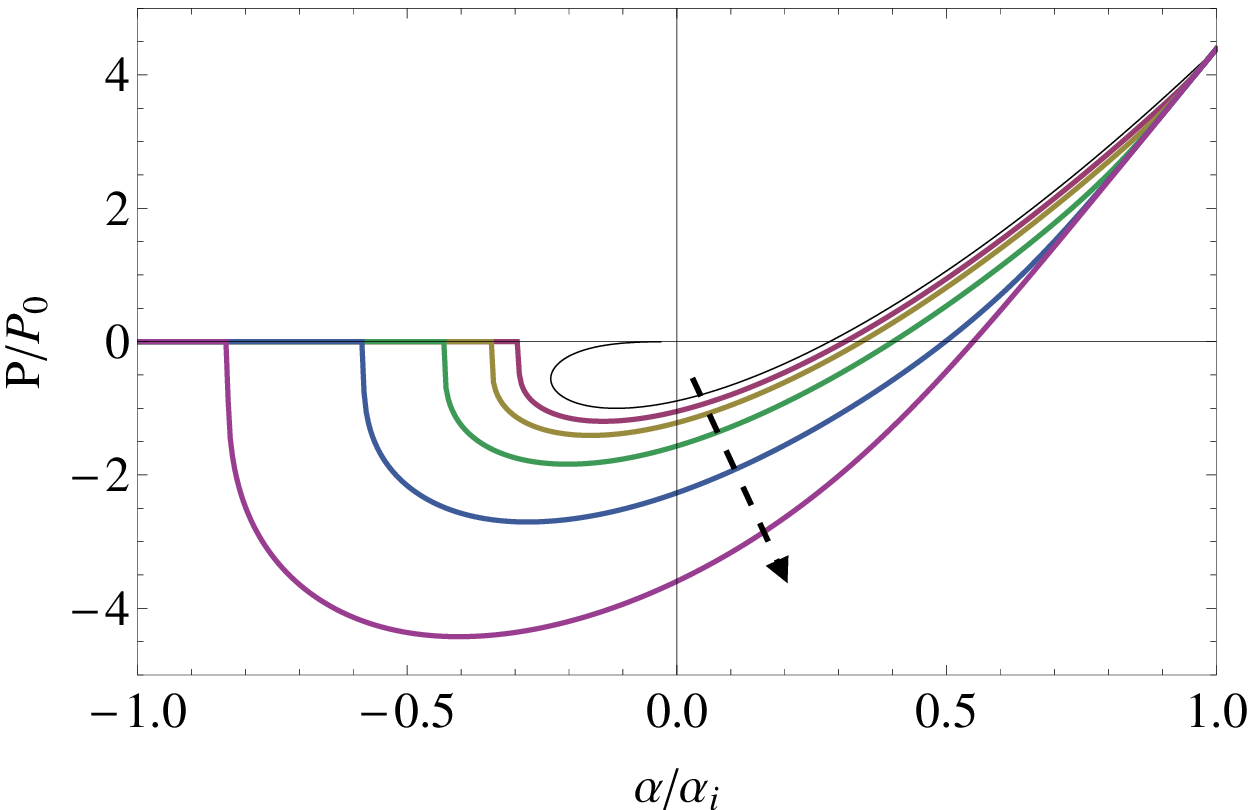}(a)
\end{tabular}

\begin{tabular}
[c]{l}%
\centering\includegraphics[height=45mm]{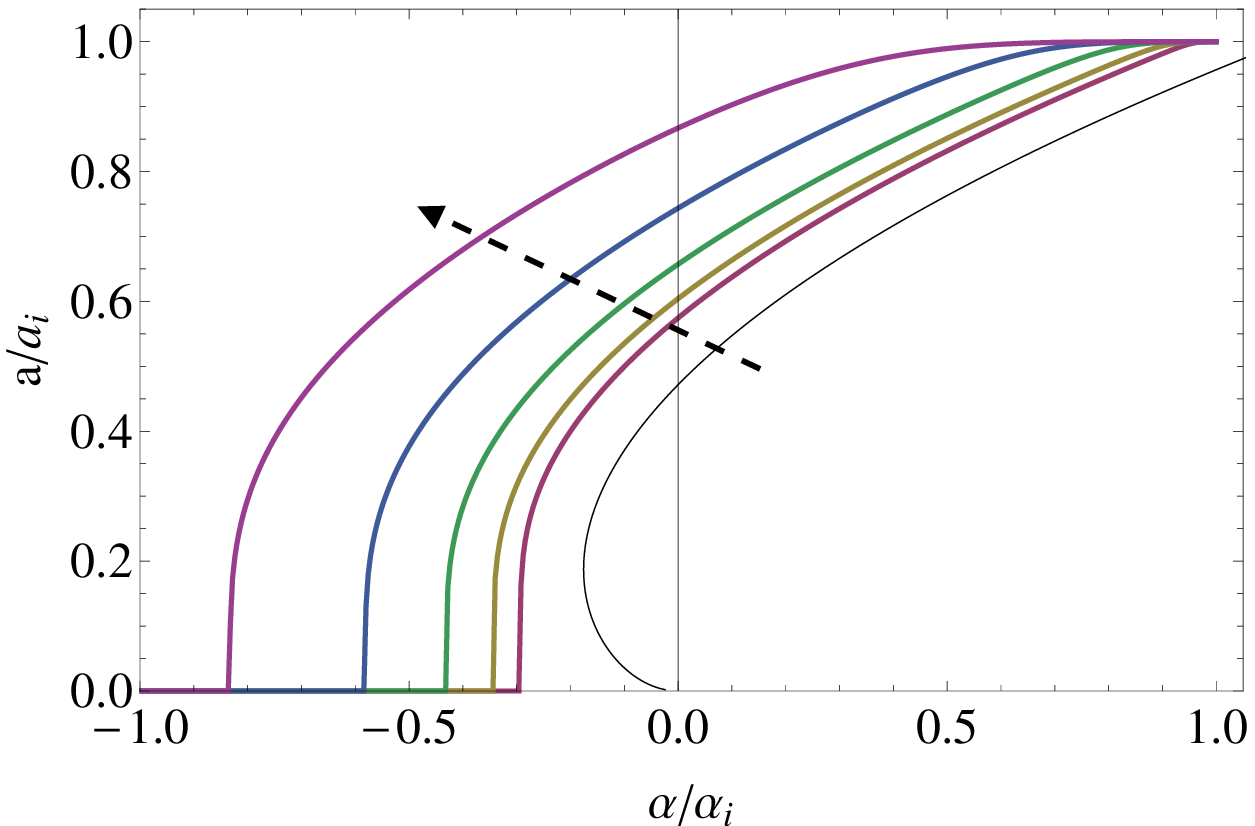}(b)
\end{tabular}

\begin{tabular}
[c]{l}%
\centering\includegraphics[height=45mm]{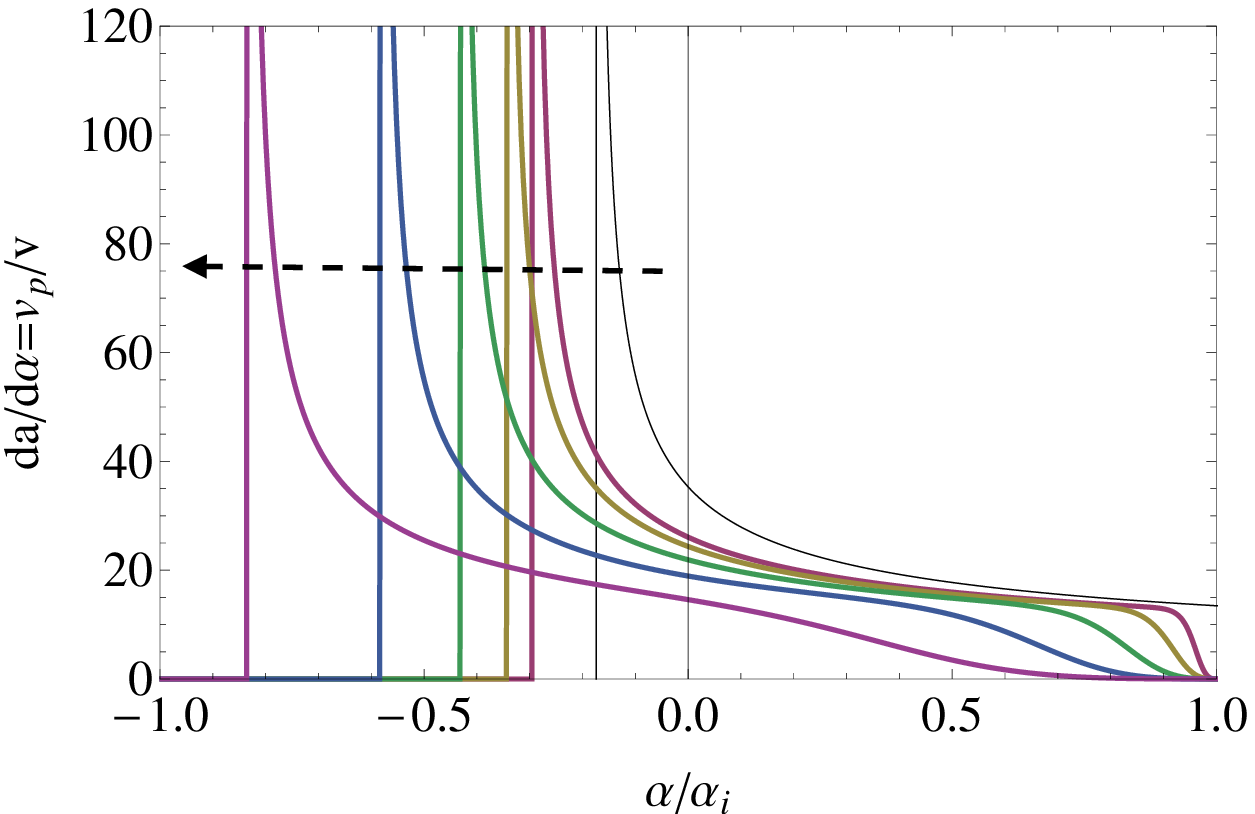}(c)
\end{tabular}

Fig.5 - The load $P/P_{0}$ (a), the contact radius $a/a_{i}$ (b), and the
velocity of contact line $da/d\alpha=v_{p}/v$ (c) as a function of approach
$\alpha/\alpha_{i}$ for the smooth sphere. The inner black curve is the JKR
classical solution, and the other 5 curves are obtained numerically for
$V=0.0002,0.002,0.02,0.2,2$ (follow the arrow). \ Here, $n=0.33$, and other
constants as indicated in Fig.1.
\end{center}

\subsection{Some comparisons}

Summarizing the pull-off results for $n=0.33$, but adding some solutions also
at different amplitudes of roughness $A$, we obtain the amplification factor
for pull-off with respect to the JKR value as in Fig.6. Notice initially that
the smooth sphere results tend to a power-law scaling (linear in the log-log
plot) as expected from the material law (\ref{wvisco}), after a transition
from the elastic behaviour. As it is evident from the figure, starting off at
low velocity with increasing amplitude of roughness, increases the "elastic"
amplification according to the Guduru theory, but eventually the effect
disappears at sufficiently large peeling speeds in the viscoelastic theory. In
other words, there seems to be a "cross-over" between the two phenomena at the
speed for which the two increases are the same. So in Fig.6, for example, for
the very "rough" spheres of $A/\lambda=0.045$ (blue line), we see that the
Guduru enhancement is larger (7.5) that the viscoelastic one in this velocity
range, and we don't see a significant decrease of its effect, but eventually
the smooth sphere result would be obtained for much larger speeds.

Obviously, with so many constants in the problem, it is not easy to give
comprehensive results, but further tests with a larger value of $n=0.6$ give
results which confirm our conclusion about the crossover: namely pull-off
force increases faster with the dimensionless speed factor $V$, but the Guduru
effect disappears also faster.

\begin{center}

$%
\begin{array}
[c]{cc}%
\begin{tabular}
[c]{l}%
\centering\includegraphics[height=45mm]{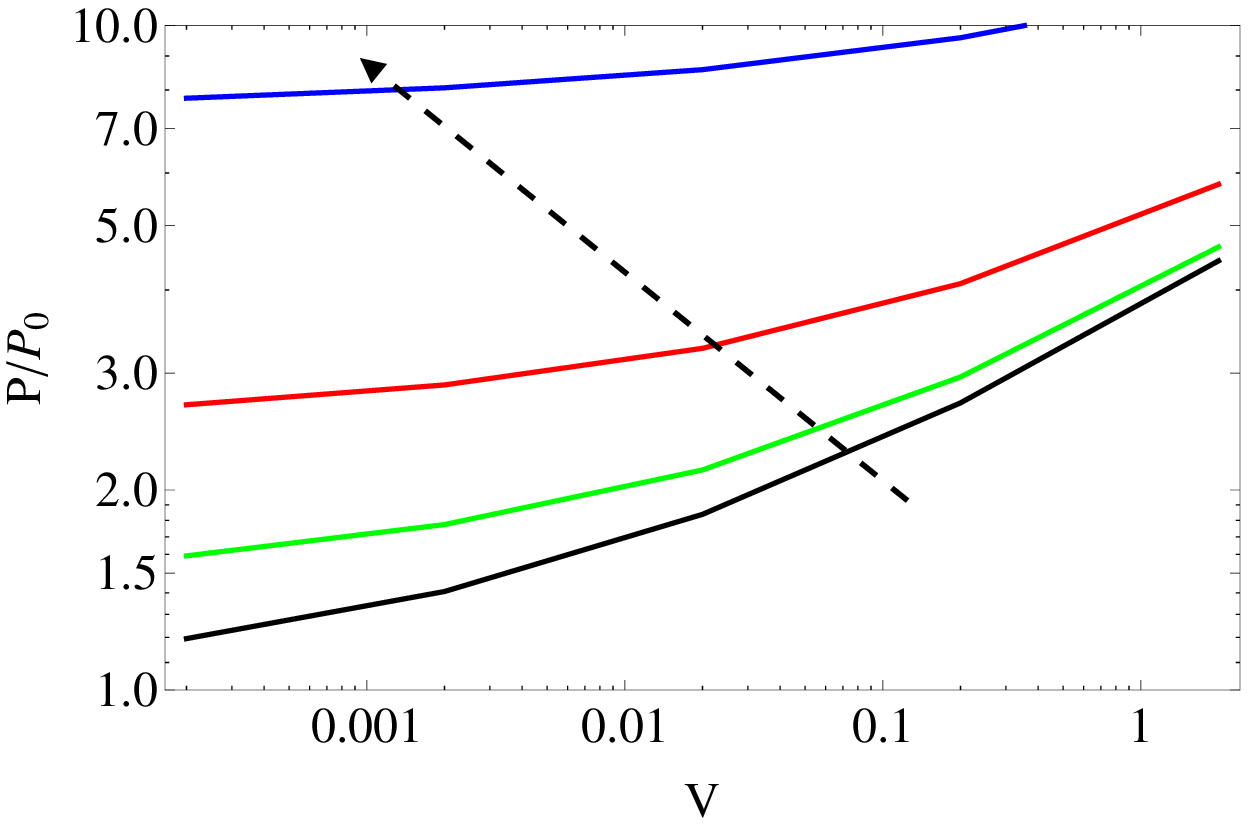}
\end{tabular}
&
\begin{tabular}
[c]{l}%
\centering\includegraphics[height=45mm]{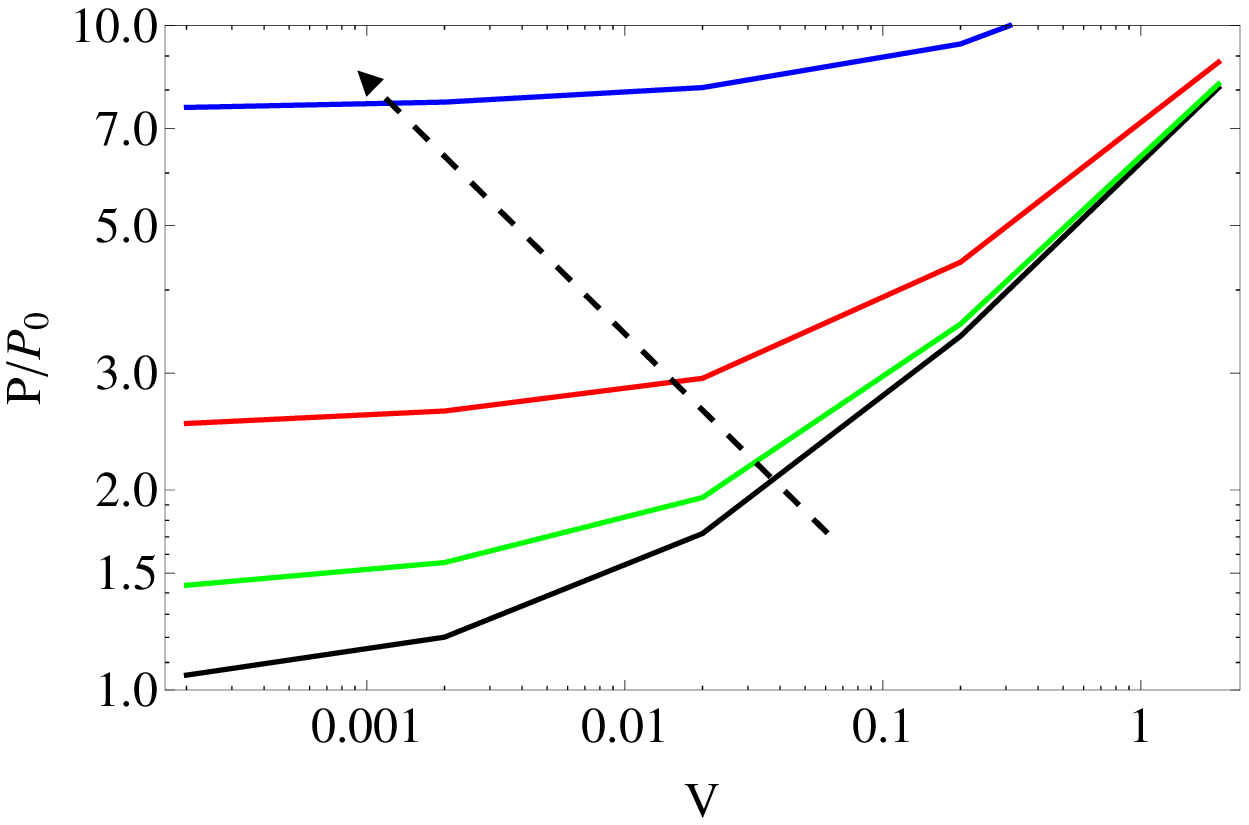}
\end{tabular}
\\
\text{(a)} & \text{(b)}%
\end{array}
$

Fig.6 - The pull-off amplification with respect to the JKR value, $P/P_{0}$ as
a function of dimensionless speed of withdrawal $V$ for various amplitude of
waviness increasing as indicated by arrow: $A/\lambda=0$ for the smooth sphere
(black), $A/\lambda=0.005$ (green)$,A/\lambda=0.015$ (red), $A/\lambda=0.045$
(blue). Here, all constants as indicated in Fig.1, except\ (a) $n=0.33$; (b)
$n=0.6$.
\end{center}

\section{Discussion}

Considering the experiments on our geometry done by Guduru \&\ Bull (2007),
their gelatin material is indeed a viscoelastic material, as recognized by
Guduru \&\ Bull (2007) who however did not characterized the material in
particular and tried to minimize the loading rate effects by keeping in their
tensile test machine a crosshead velocity at $v=3mm/min=50\mu m/s$ in all
experiments. Indeed, even the smooth sphere case they use for measuring the
baseline work of adhesion shows significant deviations between loading and the
unloading curves which are not present in JKR theory. The unloading has
specific features similar to what we found for the smooth sphere, namely when
unloading begins, the contact radius does not begin to decrease
immediately.\ Fitting JKR curves, Guduru \&\ Bull (2007) extracted
$w_{0}=0.008N/m$ during loading and $w_{0}^{\prime}=0.22N/m$, a difference of
a factor $27.\,\allowbreak5$ but they used as baseline for their comparison
with the wavy surfaces the unloading value. Despite experimental results
capture generally the trend of the predictions, there is a "systematic
difference between the experimental observation and the theoretical
prediction" as the authors say, of the order of a $-25\%$. \ We attribute this
to the effect observed in the present paper, namely that since Guduru \&\ Bull
(2007) used their elastic theory with the work of adhesion already increased
by viscoelastic effects, they overestimate the effect of load amplification.

\bigskip Consider next the experiments by Barquins \& Maugis (1981), where
material was characterized: the viscoelastic toughness (\ref{wvisco}) was
measured with $n=0.6$ in the range $v_{p}=10^{-1}-10^{3}\mu m/s$, i.e. over 4
decades of speed, which corresponds to an increase of 2 orders of magnitude in
$\frac{w-w_{0}}{w_{0}}$\footnote{Measuring $v_{p}\left[  \mu m/s\right]  $, we
have $\frac{w-w_{0}}{w_{0}}\simeq10v_{p}^{0.6}$from $w\simeq3.5w_{0}$ to
$w\simeq632w_{0}$.}. In "tack" experiment of a $2.2mm$ glass ball on the
polyurethane surface (i.e. pull-off experiments for which after a fixed
contact time (5 min), at room temperature, a cross-head velocity is imposed in
a tensile machine), an adherence force more than 30 times the quasistatic
value at fixed displacement of the JKR theory, was found when $v=1\mu m/s$. By
increasing the cross-head speed by 4 orders of magnitude i.e. to $v=$1$mm/s$,
pull-off increased by a further factor of about 30. That is the viscoelastic
effect can be very large, and therefore generally much larger than the Guduru
effect. Also, Guduru effect only holds for a quite special waviness, and when
the contact "peels" quite uniformly around a circle, and requires the initial
contact area to be compact, which poses some limits to the amplitude of
roughness (see a more general numerical solution using Lennard-Jones
force-separation law in Papangelo \& Ciavarella, 2020). Further, Li \textit{et
al}. (2019) numerical experiments for a rough sphere and elastic materials but
with 2-dimensional wavy roughness (a generalization of the Guduru geometry
which does not require the contact to peel in axisymmetric manner), the
adhesion enhancement still persists although perhaps reduced (they found an
increase of a factor 1.7 for a Johnson parameter (\ref{alfaKLJ}) $\alpha
_{KLJ}=0.37$ which corresponds to a Guduru 1D enhancement (\ref{enhancement})
of \bigskip\
\begin{equation}
\frac{w_{eff}}{w_{0}}=\left(  1+\frac{1}{\sqrt{\pi}0.37}\right)  ^{2}=6.37
\end{equation}
and therefore the 2D sinusoidal waviness already decreases the Guduru effect
significantly. Further numerical experiments in Li \textit{et al}. (2019) with
random roughness suggest that this enhancement was reduced to a maximum factor
of 1.2, much smaller than expected from Guduru theory because of the random
nature of roughness.

Returning finally to the experiments of Dalvi \textit{et al.} (2019), the
increase of apparent work of adhesion in the smoother specimen (Polished
UltraNanoCrystalline Diamond (PUNCD)) is of a factor 2, at the retraction
speed of $60nm/s$. Given the discussed experimental findings of Barquins \&
Maugis (1981), which we can compare only qualitatively as the material may be
quite different, if a factor 30 can be justified at $v=1\mu m/s$, the factor 2
increase cannot be excluded not even at the $60nm/s$ speed. Regarding the
roughness-induced enhancement, it was of a further factor less than 2, which
could be either due to area increase as from the Persson-Tosatti (2001)
theory, or from a reduced Guduru effect. But in view of the fact that the two
enhancements are of the same order, this factor 2 may be too large, according
to the present results and considering, to be attributed to Guduru's effect,
especially given the roughness is probably of random nature, and so assuming
Li \textit{et al}. (2019)'s results to be appropriate.

\section{Conclusions}

We have revisited the Guduru model for roughness-induced enhancement of
adhesion of a sphere/flat contact, adding the effect of viscoelasticity which
is expected in soft materials. The results have demonstrated that the
roughness-induced amplification of pull-off in the Guduru model, which
effectively can be modelled as an increased work of adhesion in the unloading
curve, is reduced progressively when velocity increases with respect to the
baseline smooth viscoelastic sphere. This is also in qualitative agreement
with the original experiments of Guduru and Bull (2007). A significant
reduction has already occurred at a "cross-over" velocity for which the two
enhancement (the Guduru and the viscoelastic one) are of equal magnitude. We
may be tempted therefore to speculate that viscoelasticity effectively damps
the roughness-induced elastic instabilities, reduces roughness effects in
unloading, while its effects are concentrated in the loading phase.

\section{Acknowledgements}

MC acknowledges support from the Italian Ministry of Education, University and
Research (MIUR) under the program "Departments of Excellence" (L.232/2016).

\section{References}

Andrews, E. H., \& Kinloch, A. J. (1974). Mechanics of elastomeric adhesion.
In Journal of Polymer Science: Polymer Symposia (Vol. 46, No. 1, pp. 1-14).
New York: Wiley Subscription Services, Inc., A Wiley Company.

Barber, M., Donley, J., \& Langer, J. S. (1989). Steady-state propagation of a
crack in a viscoelastic strip. \textit{Physical Review A}, 40(1), 366.

Barquins, M., \& Maugis, D. (1981). Tackiness of elastomers. \textit{The
Journal of Adhesion}, 13(1), 53-65.).

\bigskip Briggs G A D and Briscoe B J, The effect of surface topography on the
adhesion of elastic solids, \textit{J. Phys. D: Appl. Phys.}1977: 10 2453--2466

Ciavarella, M. (2016). On roughness-induced adhesion enhancement. \textit{The
Journal of Strain Analysis for Engineering Design,} 51(7), 473-481.

Dahlquist, C. A. in Treatise on Adhesion and Adhesives, R. L. Patrick (ed.),
Dekker, New York, (1969a), 2, 219.

Dahlquist, C., Tack, in Adhesion Fundamentals and Practice. (1969b), Gordon
and Breach: New York. p. 143-151.

Dalvi, S., Gujrati, A., Khanal, S. R., Pastewka, L., Dhinojwala, A., \&
Jacobs, T. D. (2019) Linking energy loss in soft adhesion to surface
roughness. Proceedings of the National Academy of Sciences, 116(51):, 25484-25490.

Fuller, K.N.G. , Roberts A.D. Rubber rolling on rough surfaces \textit{J.
Phys. D Appl. Phys.} 1981: 14, 221--239

Fuller, K. N. G., \& Tabor, D. (1975). The effect of surface roughness on the
adhesion of elastic solids. \textit{Proc Roy Soc London A}; 345:1642, 327-342

Kendall, K. \ Molecular adhesion and its applications: the sticky universe,
Springer Science and Business Media (2007).

Gent, A. N., \& Schultz, J. (1972). Effect of wetting liquids on the strength
of adhesion of viscoelastic material. \textit{The Journal of Adhesion}, 3(4), 281-294.

Gent, A. N., \& Petrich, R. P. (1969). Adhesion of viscoelastic materials to
rigid substrates. \textit{Proceedings of the Royal Society of London. A.
Mathematical and Physical Sciences}, 310(1502), 433-448.

Guduru, P.R. Detachment of a rigid solid from an elastic wavy surface: theory
\textit{J. Mech. Phys. Solids: }2007; 55, 473--488

Guduru, P.R. , Bull, C. Detachment of a rigid solid from an elastic wavy
surface: experiments \textit{J. Mech. Phys. Solids}, 2007: 55, 473--488

\bigskip Kesari, H., \& Lew, A. J. Effective macroscopic adhesive contact
behavior induced by small surface roughness. \textit{Journal of the Mechanics
and Physics of Solids},2011: 59(12), 2488-2510.

\bigskip

Greenwood, J. A., \& Johnson, K. L. (1981). The mechanics of adhesion of
viscoelastic solids. \textit{Philosophical Magazine A}, 43(3), 697-711.

Johnson, K.L. , Kendall, K. , Roberts, A.D., (1971) Surface energy and the
contact of elastic solids. \textit{Proc R Soc Lond} : A324:301--313. doi: 10.1098/rspa.1971.0141

Johnson, K. L. (1995). The adhesion of two elastic bodies with slightly wavy
surfaces. International Journal of Solids and Structures, 32(3-4), 423-430.

Li, Q., Pohrt, R., \& Popov, V. L. Adhesive Strength of Contacts of Rough
Spheres. (2019) \textit{Frontiers in Mechanical Engineering}, 5: 7.

Maugis, D., \& Barquins, M. Fracture mechanics and adherence of viscoelastic
solids. In: Adhesion and adsorption of polymers. Springer, Boston, MA, 1980.
p. 203-277.

Muller, V. M. (1999). On the theory of pull-off of a viscoelastic sphere from
a flat surface. \textit{Journal of Adhesion Science and Technology}, 13(9), 999-1016

Papangelo, A., \& Ciavarella, M. (2020) A Numerical Study on Roughness-Induced
Adhesion Enhancement in a Sphere with an Axisymmetric Sinusoidal Waviness
Using Lennard--Jones Interaction Law. \textit{Lubricants}, 8(9): 90.

Persson, B. N. J., \& Brener, E. A. (2005). Crack propagation in viscoelastic
solids. \textit{Physical Review E}, 71(3), 036123.

\bigskip

Persson, B. N. J., \& Tosatti, E. (2001) The effect of surface roughness on
the adhesion of elastic solids. \textit{The Journal of Chemical Physics},
115(12): 5597-5610

Roberts, A. D. (1979). Looking at rubber adhesion. \textit{Rubber Chemistry
and Technology}, 52(1), 23-42.

Rodriguez, N., Mangiagalli, P., \& Persson, B. N. J. (2020). Viscoelastic
crack propagation: review of theories and applications. arXiv preprint arXiv:2009.04936.

Tiwari, A., Dorogin, L., Bennett, A. I., Schulze, K. D., Sawyer, W. G., Tahir,
M., ... \& Persson, B. N. J. (2017). The effect of surface roughness and
viscoelasticity on rubber adhesion. \textit{Soft matter}, 13(19), 3602-3621.

\bigskip

Williams, M. L.; Landel, R. F.; Ferry, J. D. The Temperature Dependence of
Relaxation Mechanisms in Amorphous Polymers and Other Glass-Forming Liquids.
\textit{Journal of the American Chemical Society }1955, 77, 3701-3707.

\end{document}